%% file: paper.tex
\documentclass[11pt]{article}
\usepackage{graphicx}

\newcommand{\BABARPubYear}    {04}

\newcommand{\BABARConfNumber} {15}
\newcommand{\SLACPubNumber} {10616}

\input babarsym
\include{mydef}

\long\def\inst#1{\par\nobreak\kern 4pt\nobreak
    {\it #1}\par\vskip 10pt plus 3pt minus 3pt}

\begin{document}
{\pagestyle{empty}
\begin{flushright}
\babar-CONF-\BABARPubYear/\BABARConfNumber \\
SLAC-PUB-\SLACPubNumber \\
%hep-ex/\LANLNumber \\
July 2004 \\
\end{flushright}

\par\vskip 5cm

% Title of the paper
\begin{center}
{\Large {\bf Measurement of the {\boldmath \Bzb} Lifetime and of the
{\boldmath \BzBzb} Oscillation Frequency Using Partially Reconstructed 
{\boldmath $\Bzb\rightarrow D^{*+} \ell^- {\overline\nu_\ell}~$} Decays}} 
\end{center}
\bigskip

\begin{center}
\large The \babar\ Collaboration\\
\mbox{ }\\
\today
\end{center}
\bigskip \bigskip

% Abstract
\begin{center}
\large \bf Abstract
\end{center}

We present a simultaneous measurement of the \Bzb\ lifetime $\tBz$ and mixing parameter \deltamd. We use a sample of about 50000  \BtoDs\ partially reconstructed 
decays identified with the \babar\ detector at the \pep2\ storage ring at SLAC, where
the flavor of the other $B$ meson is determined from the charge of another high momentum 
lepton in the same event. The preliminary results are
\ba
\nonumber \tBz &=& (1.501  \pm 0.008  ~\mathrm{(stat.)} ~\pm0.030  ~\mathrm{(syst.))~ps}, \\
\nonumber \dmd &=& (0.523  \pm 0.004  ~\mathrm{(stat.)} ~\pm0.007  ~\mathrm{(syst.))~ps}^{-1}.
\ea

\vfill
\begin{center}

Submitted to the 32$^{\rm nd}$ International Conference on High-Energy Physics, ICHEP 04,\\
16 August---22 August 2004, Beijing, China

\end{center}

\vspace{1.0cm}
\begin{center}
{\em Stanford Linear Accelerator Center, Stanford University, 
Stanford, CA 94309} \\ \vspace{0.1cm}\hrule\vspace{0.1cm}
Work supported in part by Department of Energy contract DE-AC03-76SF00515.
\end{center}

\newpage
} % end of pagestyle{empty}

% Input author list file
\input authors_sum2004.tex

% The body of the paper starts here
\section{INTRODUCTION}
\label{sec:Introduction}
\input{Intro}

\section{THE \babar\ DETECTOR AND DATASET}
\label{sec:babar}
\input{Dete}

\section{ANALYSIS METHOD}
\label{sec:Analysis}
\input{Ana}

\section{RESULTS}
\label{sec:Physics}
\input{Res}

\section{SYSTEMATIC UNCERTAINTIES}\label{sec:syst}
\label{sec:Systematics}
\input{Syst}

\section{CONCLUSION}
\label{sec:Summary}
\input{Conclusion}

\section{ACKNOWLEDGMENTS}
\label{sec:Acknowledgments}

\input acknowledgements

\end{document}

%% file: mydef.tex
\newcommand{\bi}{\begin{itemize}}
\newcommand{\ei}{\end{itemize}}
\newcommand{\ben}{\begin{enumerate}}
\newcommand{\een}{\end{enumerate}}
\newcommand{\bc}{\begin{center}}
\newcommand{\ec}{\end{center}}
\newcommand{\bt}{\begin{table}}
\newcommand{\et}{\end{table}}
\newcommand{\be}{\begin{equation}}
\newcommand{\eeq}{\end{equation}}
\newcommand{\ba}{\begin{eqnarray}}
\newcommand{\ea}{\end{eqnarray}}
\newcommand{\hs}{\hspace}

\newcommand{\la}{\ifmmode {\leftarrow} \else {$\leftarrow$}\fi}
\newcommand{\Ra}{\ifmmode {\Rightarrow} \else {$\Rightarrow$}\fi}
\newcommand{\La}{\ifmmode {\Leftarrow} \else {$\Leftarrow$}\fi}
\newcommand{\Lra}{\ifmmode {\Longrightarrow} \else {$\Longrightarrow$}\fi}
\newcommand{\Lla}{\ifmmode {\Longleftarrow} \else {$\Longleftarrow$\fi}}
\newcommand{\Llra}{\ifmmode {\Longleftrightarrow} \else {$\Longleftrightarrow$\fi}}
\newcommand{\Lk}{\ifmmode {{\cal L}} \else {${\cal L}$}\fi}
\newcommand{\Wt}{\ifmmode {{\cal W}} \else {${\cal W}$}\fi}
\newcommand{\Br}{\ifmmode {{\cal B}} \else {${\cal B}$}\fi}
\newcommand{\N}{\ifmmode {{\cal N}} \else {${\cal N}$}\fi}
\newcommand{\G}{\ifmmode {{\cal G}} \else {${\cal G}$}\fi}
\newcommand{\E}{\ifmmode {{\cal E}} \else {${\cal E}$}\fi}
\newcommand{\tBz}{\ifmmode {\tau_{\Bz}} \else { $\tau_{\Bz}$ } \fi }
\newcommand{\tBu}{\ifmmode {\tau_{\Bub}} \else { $\tau_{\Bub}$ } \fi }
\newcommand{\BtoDs}{\mbox{$\Bbar^0\rightarrow D^{*+} \ell^- \bar{\nu_\ell}$}}
\newcommand{\BtoDss}{\mbox{$B\rightarrow D^{*} \pi \ell^- \bar{\nu_\ell}$}}

\newcommand{\pstar}{\ifmmode {\pi^{+}_{soft}} \else {$\pi^{+}_{soft}$} \fi }
\newcommand{\psoft}{\ifmmode {\pi^{+}_{s}} \else {$\pi^{+}_{s}$} \fi }
\newcommand{\plab}{\ifmmode{p} \else {$p$} \fi}
\newcommand{\ks}{\ifmmode{k^*} \else {$k^*$} \fi}
\newcommand{\mnusq}{\ifmmode{{{\cal M}^2_\nu}} \else {${{\cal M}_{\nu}}^2$}\fi} 
\newcommand{\DTau}{\ifmmode {\Delta \tau} \else {$\Delta \tau$}\fi}
\newcommand{\dmd}{\ifmmode {\Delta m_d} \else {$\Delta m_d$}\fi}
\newcommand{\deltams}{\ifmmode {\Delta m_s} \else {$\Delta m_s$}\fi}
\newcommand{\ggcc}{\ifmmode {GeV^2/c^4} \else {$GeV^2/c^4$}\fi}
\def\BpBm {\ensuremath{B^+ {\kern -0.16em \Bub}}}

\def\ptl{\ensuremath{{\cal P}\ell}}
\def\ctl{\ensuremath{{\cal C}\ell}}
\def\cdl{\ensuremath{{\cal D}\ell}}

\def\tDe{\ensuremath{\tau_{D_e}}}
\def\dz{\ensuremath{\Delta z}}

\def\st{\ensuremath{\sigma_{\Delta t}}}

\def\chid{\ensuremath{\chi_d}}

%% file: authors_sum2004.tex
\begin{center}
\small

The \babar\ Collaboration,
\bigskip

%% author list as of 02-Jul-2004 (609 authors)
%
B.~Aubert,
R.~Barate,
D.~Boutigny,
F.~Couderc,
J.-M.~Gaillard,
A.~Hicheur,
Y.~Karyotakis,
J.~P.~Lees,
V.~Tisserand,
A.~Zghiche
\inst{Laboratoire de Physique des Particules, F-74941 Annecy-le-Vieux, France }
% ...

\end{center}\newpage

%% file: Intro.tex
The time evolution of \Bzb\ mesons 
is governed by the overall decay
rate $\Gamma(\Bzb) = 1/\tBz$ and by the mass difference \deltamd\ of the two mass
eigenstates.
A precise determination of $\Gamma(\Bzb)$ reduces the systematic error
on the parameters \Vcb\ and \Vub\ of the Cabibbo-Kobayashi-Maskawa flavor mixing matrix.
The parameter $|V_{td}V_{tb}^*|$ enters the box diagram that is responsible for \BzBzb\ oscillations and
can be determined from a measurement of \deltamd, with a sizeable systematic error due to
theoretical uncertainties. 
\par
We describe here a measurement of $\tBz$ and \deltamd\ performed using
\BtoDs\ decays\footnote{Charge conjugate states are always implicitly assumed; $\ell$ means either electron
or muon.} 
selected from a sample of about 88 million \BB\ events recorded by the \babar\ 
detector at the \pep2\ asymmetric-energy \epem\ storage ring, operated at or near the
\FourS\ resonance. \BB\ pairs from the \FourS\ decay move along the beam axis with a nominal 
Lorentz boost $\langle \beta\gamma \rangle = 0.55$, so that the vertices from the two $B$-decay points are
separated on average by about 260 $\mu$m.
The \BzBzb\ system is produced in a coherent $P$-wave state, so that flavor oscillation is
measurable only relative to the decay of the first $B$ meson. Mixed (unmixed) events are selected 
by the observation of two equal (opposite) flavor $B$ meson decays.
The probabilities of observing
mixed (${\cal S}^{-}$) or unmixed (${\cal S}^{+}$) events as a function of the proper time 
difference \deltat\ are
\ba
\label{eq:pdf} {\cal S^{\pm}} = \frac{e^{-|\deltat|/\tBz}}{4\tBz} (1 \pm {\cal D} \cos(\deltamd \deltat) ),
\ea
where ${\cal D}$ is related to the fraction $w$ of events with wrong 
flavor assignment by the relation  ${\cal D}=1-2w$ and \deltat\ is computed from the distance between
the two vertices projected along the beam direction.
\par

%% file: Dete.tex
We have analyzed a data sample of 81 \invfb\ collected by \babar\ on the \FourS\ resonance, a sample of
9.6 \invfb\ below the resonance, to study the continuum background, 
and a sample of \BB\ simulated events corresponding to about three times 
the size of the data sample. 
The simulated events are processed through the same analysis 
chain as the real data.
\babar\ is a multi-purpose detector, described in detail in Ref.~\cite{ref:babar}. The momentum of
charged particles is measured by 
the tracking system, which consists of a silicon vertex tracker (SVT) 
and a drift chamber (DCH) in a 1.5-T magnetic field.
The SVT measures
the momentum of low transverse momentum charged tracks that do not reach the DCH due to 
bending in the magnetic field. 
The energy loss in the SVT is used in this analysis to discriminate low-momentum
pions from electrons. 
Higher-energy electrons are identified from the ratio of the energy of their associated shower in 
the electromagnetic calorimeter (EMC) to their momentum, the transverse profile of the 
shower, the energy loss in the DCH, and the information from the Cherenkov detector 
(DIRC). The electron identification efficiency within the tracking volume is about $90\%$, and 
the hadron misidentification probability is less than $1\%$. 
Muons are identified on the basis of the energy deposit
in the EMC and the penetration in the instrumented flux return. Muon candidates
compatible with the kaon hypothesis in the DIRC are rejected. The muon identification 
efficiency is about $60\%$, and the hadron misidentification rate is about 2\%.
\par

%% file: Ana.tex
\subsection{Selection of \BtoDs\ decays}

We select events that have more than four charged tracks.
We reduce the contamination from light-quark production in continuum events by
requiring the normalized Fox-Wolfram second moment~\cite{ref:FW} to be less than 0.5.
We select \BtoDs\ events with partial reconstruction of the decay $\dsp \ra \psoft \Dz$,
using only the charged lepton and the soft pion (\psoft) from the \dsp .
The \Dz\ is not reconstructed, resulting in high reconstruction efficiency.
\babar\ has already published two measurements of $\tBz$ \cite{ref:t1,ref:t2} and a measurement of 
$\sin(2\beta + \gamma)$~\cite{ref:sin2bg} based on partial 
reconstruction. This technique
was originally applied to \BtoDs\ decays by ARGUS \cite{ref:ARGUS}, and then used
by CLEO \cite{ref:CLEO}, DELPHI \cite{ref:DELPHI}, and OPAL \cite{ref:OPAL}. This is, however, the
first simultaneous measurement of $\tBz$ and \deltamd\ based on partial reconstruction.
\par
\begin{figure}[!htb]
\begin{center}
\includegraphics[width=12cm,height=14cm]{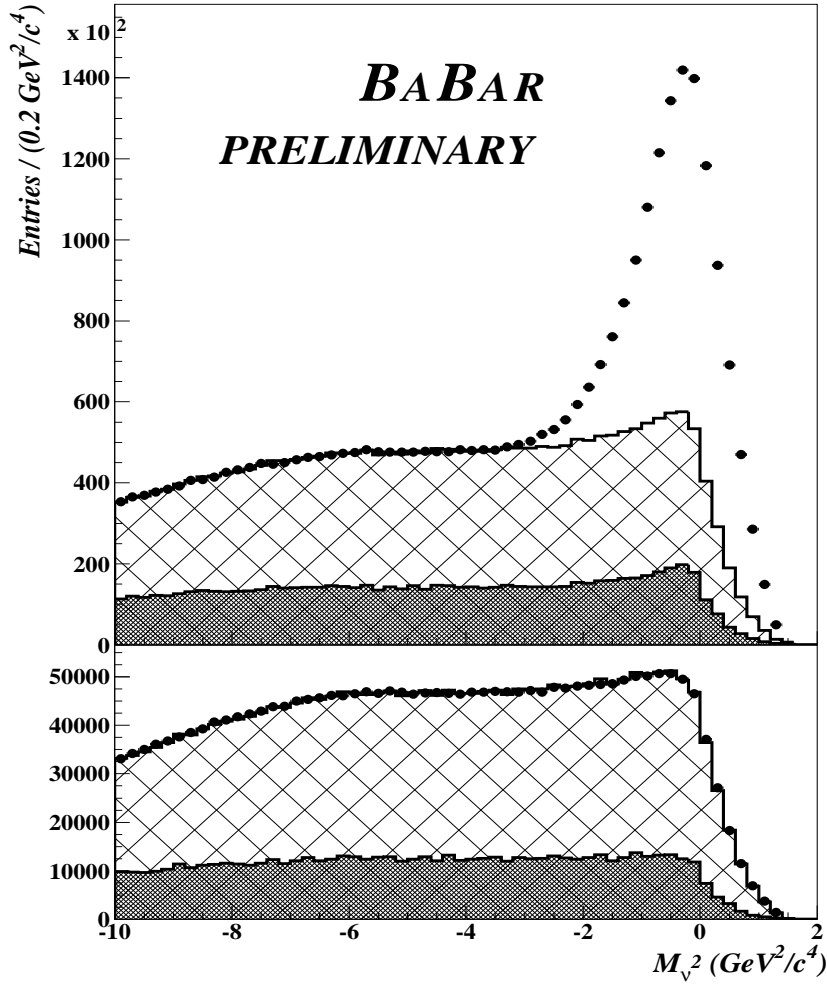}
\end{center}
\caption{$\mnusq$ distribution for the right-charge (top) and wrong-charge (bottom) events.
The points correspond to on resonance data. The distributions of continuum events 
(dark histogram), obtained from rescaled off-peak events, and \BB\ combinatorial events (hatched area), 
obtained from the simulation, are overlaid.}
\label{f:mnu}
\end{figure}

To suppress leptons from several background sources, we use only high momentum
leptons, in the range $ 1.3 < p < 2.4 $ GeV/$c$.\footnote{The lepton and pion momenta, the \psoft direction, and $\tilde{E}_\dsp$ (see below) are always computed in the \FourS\ rest frame.}
The \psoft\ candidates have momenta 
between 60 and 200 \mevc. Due to the limited phase space available in the \dsp\ decay,
the \psoft\ is emitted within approximately a one-radian full opening angle
cone centered about the \dsp\ flight direction (in
the \FourS\ frame). Therefore, we approximate the direction of the \dsp\ to be that of the \psoft
and estimate the energy  $\tilde{E}_{D^{*+}}$ of the \dsp\ 
as a function of the energy of the \psoft\ 
using a third order polynomial, with parameters taken from the simulation.
We define the square of the missing neutrino mass as
\ba
\nonumber \mnusq = ( \frac{\sqrt{s}}{2} - \tilde{E}_{\dsp} - E_\ell )^2 - 
(\tilde{\bf{p}}_{\dsp} + {\bf{p}}_\ell )^2 ,
\ea
where we neglect the momentum of the \Bzb\ in the \FourS\ frame (on average, 0.34 GeV/$c$), 
and identify its energy with the beam energy $\sqrt{s}/2$ in the \epem\ center-of-mass frame. 
$E_\ell$ and ${\bf{p}}_\ell$ are the energy and momentum vector of the lepton and 
$\tilde{\bf{p}}_{\dsp}$ is the estimated momentum vector of the \dsp. 
The distribution of $\mnusq$ peaks at zero for signal events, while it is spread over a
wide range for background events (see Fig.~\ref{f:mnu}).\par
We determine the \Bzb\ decay point from a vertex fit of the $\ell$ and \psoft\ tracks,
constrained to the beam spot position in the plane perpendicular to the beam axis
(the $x-y$ plane). The beam spot position and size are determined on a run-by-run basis using two-prong events
\cite{ref:babar}. Its size in the horizontal ($x$) direction is 120 $\mu$m. Although the beam
spot size in the vertical ($y$) direction is only 5.6 $\mu$m, we use a constraint of 50 $\mu$m in the vertex fit
to account for the flight distance of the \Bzb\ in the $x-y$ plane. We reject events
for which the $\chi^2$ probability of the vertex fit, ${\cal P}_V$, does not exceed $0.1\%$.\par
We then apply a selection criterion to a combined signal likelihood, calculated from $p_\ell$, 
$p_\psoft$, and  
${\cal P}_V$, which results in a signal-to-background ratio of about one-to-one in the
signal region defined as $\mnusq > -2.5$ GeV$^2/c^4$. Figure \ref{f:mnu} shows the distribution
of $\mnusq$ for the events selected at this stage of the analysis. The plot on the
top is obtained from the events in which the $\ell$ and the $\pi_s$ have opposite 
charges (``right-charge''), the plot at the bottom from events where $\ell$ and the $\pi_s$ have 
equal charges (``wrong-charge''). We use wrong-charge events as a background control sample, 
to verify that the \BB\ combinatorial background is described well by the simulation. To do this,
we add the off-peak events, scaled by the ratio of the on-peak to the off-peak luminosities,
to the \BB\ Monte Carlo 
scaled to match the data in the  
region $\mnusq < -4.5$ GeV$^2/c^4$. 
We then compare the expected number of events to the number of data events in the signal region.
This ratio is $0.996 \pm 0.002$, consistent with unity. 
For the rest of the analysis we only consider right-charge events.

\subsection{Tag Vertex and B Flavor Tagging}

We restrict the analysis to events in which another charged lepton (``tagging lepton'') is found.
To reduce contamination from fake leptons and leptons originating from charm decays, we require that 
the momentum of the lepton exceeds 1.0 GeV/$c$ for electrons, 1.1 GeV/$c$ for muons.
The tagging lepton is used to tag the flavor of the other (``tag'') $B$. The decay point of the tag $B$ 
is computed 
from the intersection of the tagging lepton track with a beam-spot ellipse  
centered on the position of the reconstructed $B$. 
The beam spot constraint is applied only in the transverse plane, and the width in the vertical 
direction is inflated to 50$\mu$m. We compute the proper time difference \deltat\ between the two 
vertices from their projected distance along the beam direction ($z$-axis), $\deltaz$,
neglecting the \Bzb\ motion in the \FourS\ rest frame (boost approximation).
We obtain $\deltat = \frac{\deltaz} { c \beta\gamma}$,
where the boost factor $\beta\gamma$ is determined from the continuously measured 
beam energies. 
To remove badly reconstructed vertices we reject all events with either $|\deltaz|>3$ mm 
or $\sigma(\deltaz) >0.5$ mm, where $\sigma(\deltaz)$ is the uncertainty on $\deltaz$,
computed for each event. 
The simulation shows that the rms of the difference between the true and measured  $\deltat$
is 0.64~ps for 70$\%$ of the events, and about 1.7~ps for the rest.
We then select only one right-charge candidate per event according to the following procedure:
if there is more than one candidate in the event, we choose the one lying in the region
$\mnusq > -2.5~$GeV$^2/c^4$. We reject the event if there is more than one candidate 
(either right or wrong charge) in this region. 
This criterion reduces by about $20\%$ the number of signal events.
For background studies, we select events in the region $\mnusq < -2.5~$GeV$^2/c^4$
if there is no candidate in the region $\mnusq > -2.5~$GeV$^2/c^4$.
We find about 50000 signal events over a background of about 27000 events in the data sample.

\subsection{Sample Composition}
\label{s:sample}
Our data sample consists of the following event types, categorized according to their origin and 
to whether or not they exhibit a peak in the $\mnusq$ distribution. 
We consider signal to be any combination of a lepton and a charged $D^*$ produced in the decay of the same 
\Bzb\ meson. Signal consists mainly of \BtoDs\ decays, with
minor contributions from $\Bzb \ra \dsp \pi^0 \ellm \nulb $,   
$\Bzb \ra \dsp \tau^- \nub_\tau $, $\Bzb \ra \dsp D_s^- $, $\Bzb \ra \dsp \overline{D} X$ with $\tau$, 
${D_s}^-$, or $\overline{D}$
decaying to an $\ellm$, and $\Bzb \ra \dsp h$, with the hadron $h$ misidentified as a muon.
A peaking \Bub\ background is due to the process $\Bub \ra \dsp \pi^- \ellm \nulb $. Non-peaking 
contributions are due to random combinations of a charged lepton candidate and a low-momentum pion candidate,
produced either in \BB\ events (\BB\ combinatorial) or in $e^+e^- \ra q\overline{q}$ interactions 
with $q=u,~d,~s$ or $c$
(continuum). We compute the sample composition separately for mixed and unmixed
events by fitting the corresponding $\mnusq$ distribution to the sum of four components: continuum, \BB\ 
combinatorial, \BtoDs\ decays, and \BtoDss\ decays. Due to the production
of one or more additional pions, these latter events have a different \mnusq\ spectrum from that of the direct process \BtoDs.
 We measure the continuum contribution from the off-resonance sample,
scaled to the luminosity of the on-resonance sample. 
We determine the $\mnusq$ distributions for the other event types from the simulation, and determine 
their relative abundance in the selected sample from a fit to the $\mnusq$ distribution for the data.
Assuming isospin conservation, we assign two thirds of \BtoDss\ decays to peaking \Bub\ background and
the rest to 
$\Bzb \ra D^* \pi \ellm \nulb $
, which we add to the signal.
We vary this fraction in the study of systematic uncertainties.
We assume $50\%$ error on the isospin hypothesis that allows
us to determine the peaking \Bub\ contribution.
\par
A possible bias in the $\mnusq$ distribution comes 
from the decay chain \mbox{$\overline{B} \ra \ellm \nulb D(X)$},~\mbox{$D\ra Y\pi^+$},
where the state Y is so heavy that the charged pion is emitted at low momentum, behaving like a \psoft.
This possibility has been extensively studied by the CLEO collaboration in Ref.~\cite{CLEODtokpi}, 
where the
three $D^+$ decay modes most likely to cause this bias
have been identified: \mbox{$\overline{K}^{*0}\omega\pi^+$}, \mbox{$K^{*-}\rho^+\pi^+$} and \mbox{$\overline{K}^{*0}\rho^0\pi^+$}. They quote a systematic error of
$\pm 2.3 \%$ on the background rate as a result of that analysis.
Based on their result, we attribute the same systematic error to the number 
of combinatorial
events below the signal mass peak.\par
\begin{figure}[!htb]
\begin{center}
\begin{tabular}{cc}
\hs{-1.5cm}\includegraphics[width=10cm,height=13cm]{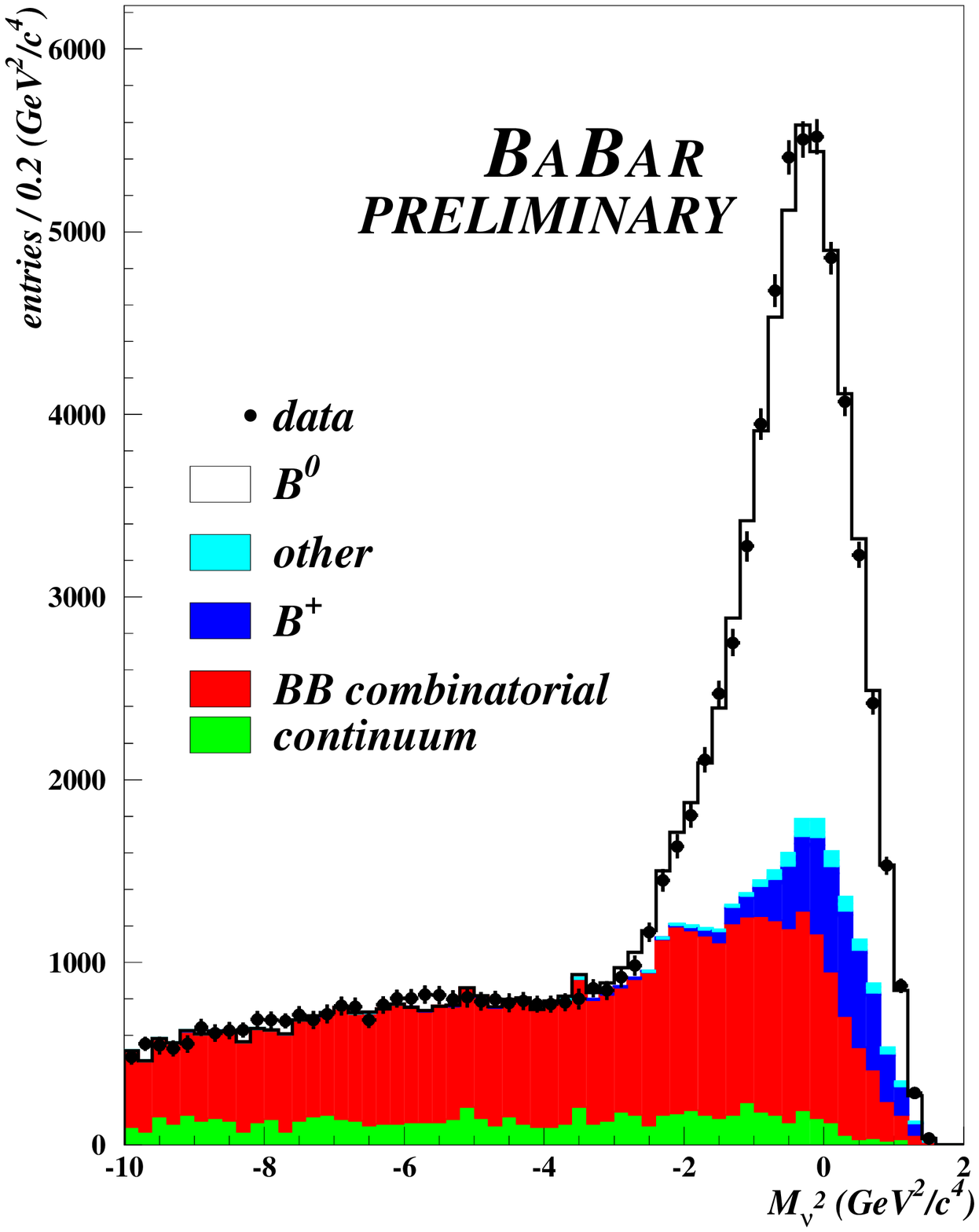} &
\hs{-1.5cm} \includegraphics[width=10cm,height=13cm]{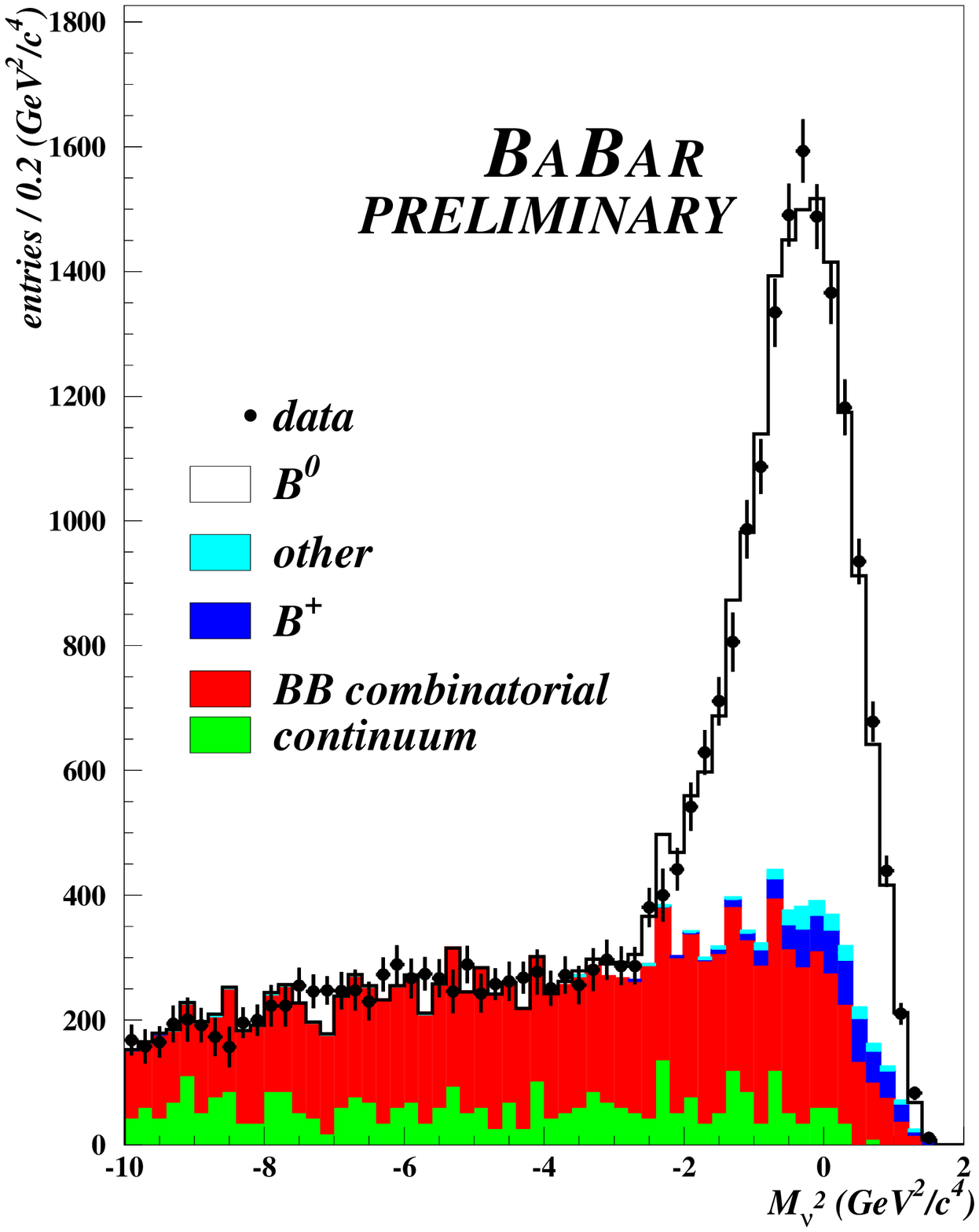}
\end{tabular} 
\end{center}
\caption{Fit to the $\mnusq$ distribution for the unmixed events 
(left) and mixed events (right).
``$\Bz$'' includes $\BtoDs$, $\Bzb \ra \dsp \pi^0 \ell \nul$,   
$\Bzb \ra \dsp \nu_\tau \tau^- $ ($\tau \ra \ell$), 
$\Bzb \ra \dsp D_s^- $ ($D_s \ra \ell$), and 
$\Bzb \ra \dsp h$ with the hadron $h$ misidentified as a muon.
``$B^+$'' includes $B^+ \ra D^{*-} \pi^+ \ell \nul$ and $B^+ \ra D^{*-} \pi^+ X$ with the $\pi^+$ misidentified as a muon. ``Other'' includes $B \ra D^* \pi \nu_\tau \tau$  ($\tau \ra \ell$), $B \ra D^* \overline{D} X $ 
($\overline{D} \ra \ell Y$), and $B \ra D^* \pi h$ with the hadron $h$ misidentified as a muon. 
 }
 \label{f:fitmnu}
\end{figure}
\begin{figure}[!htb]
\begin{center}
\begin{tabular}{cc}
\hs{-1cm}\includegraphics[width=9cm,height=13.5cm]{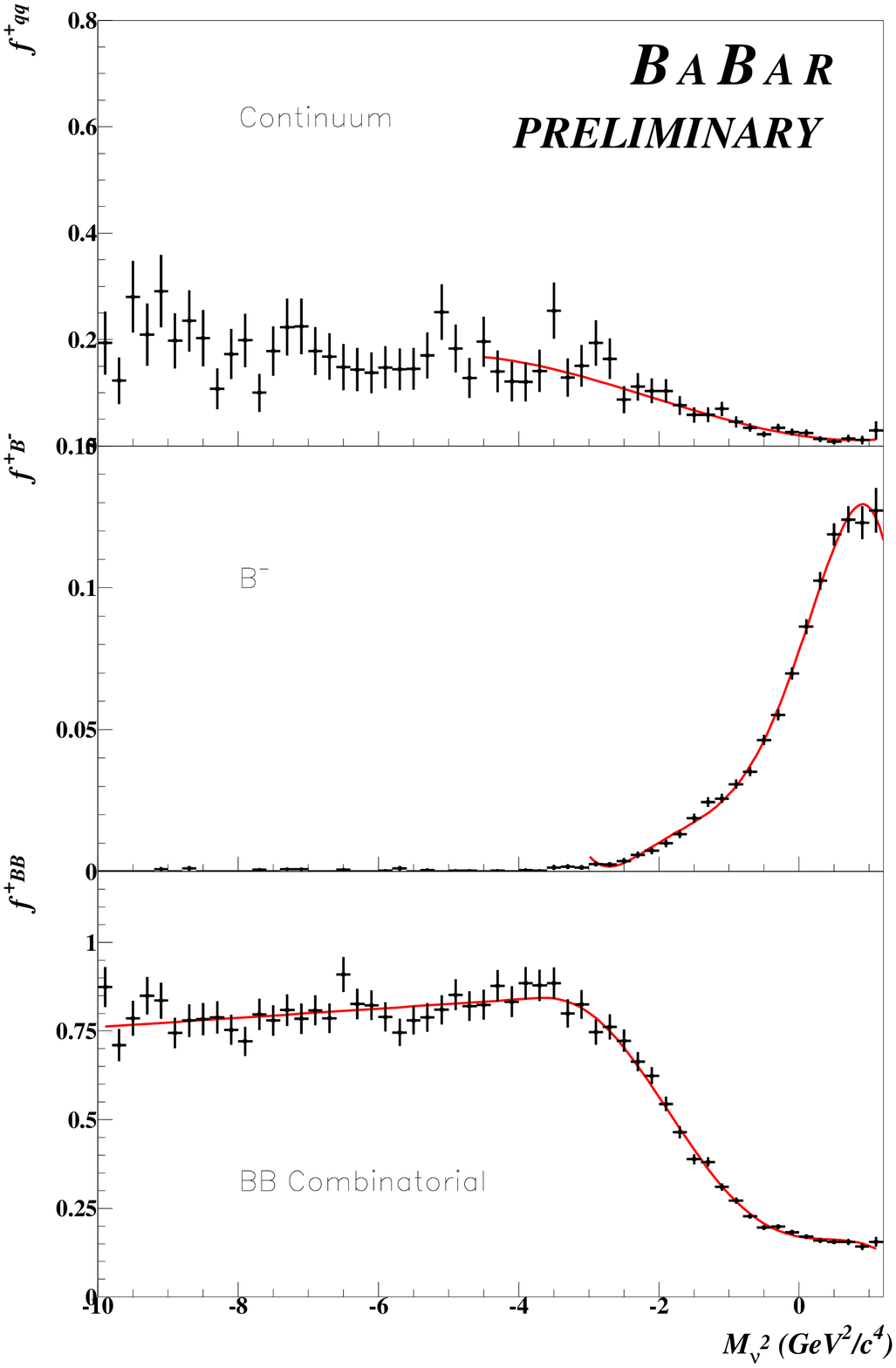} &
\hs{-1cm}\includegraphics[width=9cm,height=13.5cm]{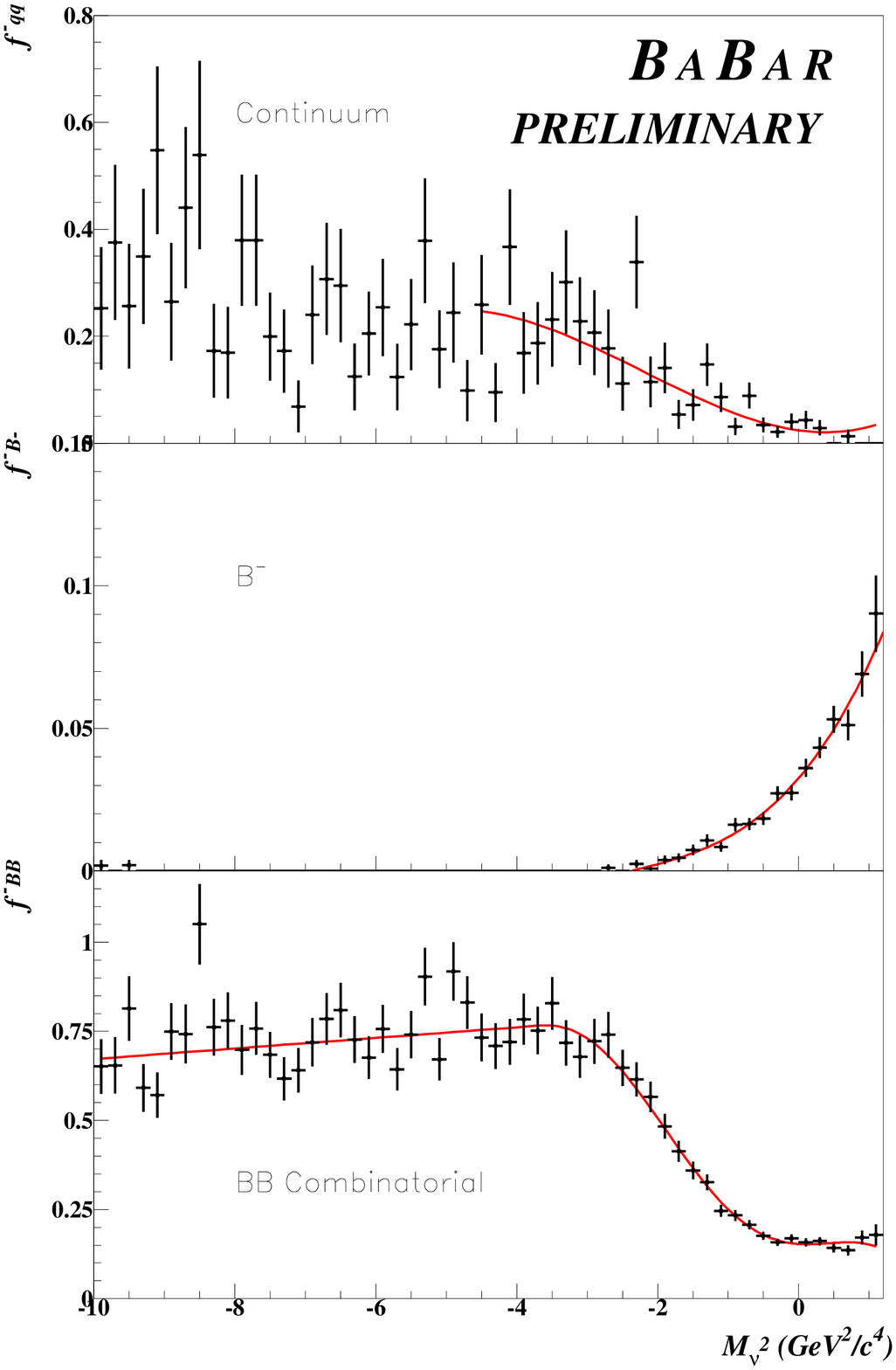}
\end{tabular} 
\end{center}
\caption{Fraction of continuum events, peaking $\Bub$, and \BB\ combinatorial background 
in the unmixed (left) and mixed (right) lepton-tagged samples. 
The fraction of continuum events, $f_{qq}^{\pm}$, is parameterized only in the 
region $\mnusq>-4.5$~GeV$^2/c^4$. For $\mnusq<-4.5$~GeV$^2/c^4$, where just continuum and combinatorial 
backgrounds are present, we assume that $f_{\Bub}^{\pm}=0$, and we compute 
$f_{qq}^{\pm}=1-f_{\BB}^{\pm}$. 
}
\label{f:frac}
\end{figure}

Figure \ref{f:fitmnu} shows the fit results for unmixed (left) and mixed (right) events.
We use the results of this study to determine the
fraction of continuum ($f_{qq}^{\pm}$), \BB\ combinatorial ($f_{\BB}^{\pm}$), and peaking \Bub\ background ($f_{\Bub}^{\pm}$)
as a function of $\mnusq$, separately for mixed ($f^-$) and unmixed ($f^+$) events. We parameterize 
these fractions with polynomial functions, as shown in Fig.~\ref{f:frac}.


\subsection{\tBz and \deltamd Determination}
We fit data and Monte Carlo events with a binned maximum-likelihood method. We divide our events into one hundred
\deltat\ bins, spanning the range $-18$~ps $< \deltat < 18$~ps, and twenty \st\ bins between 0 and 3 ps.
We assign to all events in each bin the values of \deltat\ and \st\ corresponding to the
center of the bin. We fit simultaneously the mixed and unmixed events. We do not use an extended likelihood,
but instead apply a constraint on the fraction of mixed events, 
which, for a sample of events with dilution ${\cal D}$, reads
\ba
\frac{{\cal N}_{mix}}{{\cal N}_{mix}+{{\cal N}_{unmix}}} = \chid \cdot {\cal D} + \frac{1-{\cal D}}{2},
\ea
where, neglecting the difference $\Delta \Gamma_d$ between the two mass eigenstates,
 the integrated mixing rate \chid\ is related to the product $x = \dmd \cdot \tBz$  by the relation
\ba
\chid = \frac{x^2}{2(1+x^2)}.
\ea 
We describe the \deltat\ distribution as the sum of the decay probabilities for signal and background events:
\ba
\label{e:SumPdf}
{\cal F}^{\pm}(\deltat,\sigma_{\deltat},\mnusq | \tBz,\deltamd) &=& 
f^\pm_{qq}(\mnusq) \cdot {\cal F}^{\pm}_{qq}(\deltat,\sigma_{\deltat}) \\ \nonumber &+& 
f^\pm_{\BB}(\mnusq) \cdot {\cal F}^{\pm}_{\BB}(\deltat,\sigma_{\deltat}) \\ \nonumber &+&
{\cal S}_{\Bub}f^\pm_{\Bub}(\mnusq)  \cdot {\cal F}^{\pm}_{\Bub}(\deltat,\sigma_{\deltat}) \\ \nonumber &+&
[1-{\cal S}_{\Bub}f^\pm_{\Bub}(\mnusq)-f^\pm_{\BB}(\mnusq)-f^\pm_{qq}(\mnusq)] \\ \nonumber &\cdot&
{\cal F}^{\pm}_{\Bzb}(\deltat,\sigma_{\deltat}|\tBz,\deltamd), 
\ea
where the functions ${{\cal F}_i^{\pm}}$ represent the probability density functions 
(PDF) for signal ($i=\Bzb$),
peaking \Bub ($i =\Bub $), \BB\ combinatorial ($i=\BB $), and continuum ($i=q\overline{q}$) events, and the 
superscript +($-$) applies to unmixed (mixed) events. 
To account for the $\pm 50\%$ uncertainty on the isospin
assumption (see Section \ref{s:sample}), 
we introduce in the PDF a global scale factor ${\cal S}_{\Bub}$ multiplying the fractions 
$f^{\pm}_{\Bub}$, 
equal for mixed and unmixed events, which we float in the fit but constrained to unity 
with 0.5 error.
\par
The signal PDF is obtained from Eq.~\ref{eq:pdf} modified to account for the finite 
resolution of the detector. The resolution function is expressed as the sum of three Gaussian 
functions, defined 
as ``narrow'', ``wide'' and ``outlier'':
\ba
{\cal R}(\delta\deltat,\sigma_{\deltat}) &=& 
  \frac{(1-f_w-f_o)}{\sqrt{2\pi} S_{n} \sigma_{\deltat}} ~
e^{-\frac{(\delta\deltat - o_{n})^2}{2 S_{n}^2 \sigma_{\deltat}^2}} + 
 \frac{f_w }{\sqrt{2\pi} S_{w} \sigma_{\deltat}} ~
e^{-\frac{(\delta\deltat - o_{w})^2}{2 S_{w}^2 \sigma_{\deltat}^2}} + 
 \frac{f_o}{\sqrt{2\pi} S_o}  ~
 e^{-\frac{(\delta\deltat - o_{o})^2}{2 S_{o}^2 }} ,
\ea
 where $\delta\deltat$ 
is the difference between the measured and the true value of \deltat , $o_{n}$ and $o_{w}$ 
are offsets, the factors $S_{n}$ and $S_{w}$
account for possible misestimation of $\sigma_{\deltat}$. The outlier term, described by 
a Gaussian function of fixed width $S_{o}$ and offset $o_{o}$, is introduced to
describe events with badly measured $\Delta t$, and accounts for less than 1$\%$ of the events.\par
We divide signal events according to the origin of the tag lepton into primary, cascade, 
and decay-side tags.
A primary tag is produced in the direct decay $\Bz \ra \ell^+ \nu_{\ell} X$. These events are described by 
Eq.~\ref{eq:pdf}, with ${\cal D}$ close to one (a small deviation from unity is expected due to hadron misidentification,
leptons from $J/\psi$, etc.). We expect small values of $o_n$ and $o_w$ for primary tags, 
because the lepton originates from the \Bz\ decay point.\par
Cascade tags, produced in the process $\Bz \ra D X, D \ra \ell Y$, are suppressed by the cut on the 
lepton momentum but still exist at a level of $9\%$ as we obtain by floating their relative abundance as
an additional parameter in the \deltamd\ fit on data.
The lepton production point is displaced from the 
\Bz\ decay point due to the finite lifetime of 
charm mesons and the \epem\ energy asymmetry. This results in a significant positive value of the offsets 
for this category.
Compared with the primary tag, the cascade lepton usually has the opposite charge correlation with the \Bz\ 
flavor. The same charge correlation is obtained when the charm 
meson is produced from the hadronization of the virtual $W$ from \Bz\ decay, usually referred to as 
``Upper Vertex'', which
results in the production of two opposite-flavor charm mesons. We account
for these facts by applying Eq.~\ref{eq:pdf} to the cascade tag events with negative 
dilution ${\cal D_{\ctl}} = -(1-2f_{UV}) = -0.65 \pm 0.08$, where we take from the PDG \cite{ref:PDG}
the ratio of the inclusive semileptonic branching ratios of the $b$ quark 
$f_{UV} = \frac{BR(\overline{b}\ra c\ra \ell^+)}{BR(\overline{b}\ra c\ra \ell^+)+BR(\overline{b}\ra \overline{c}\ra \ell^-)}= 0.17\pm0.04$ (the contribution to the dilution from other sources associated with the
$\psoft \ell$ candidate, such as fake hadrons, is
negligible).\par
\begin{figure}[!htb]
\begin{center}
\begin{tabular}{ccc}
\hs{-1.5cm}\includegraphics[width=8cm,height=9cm]{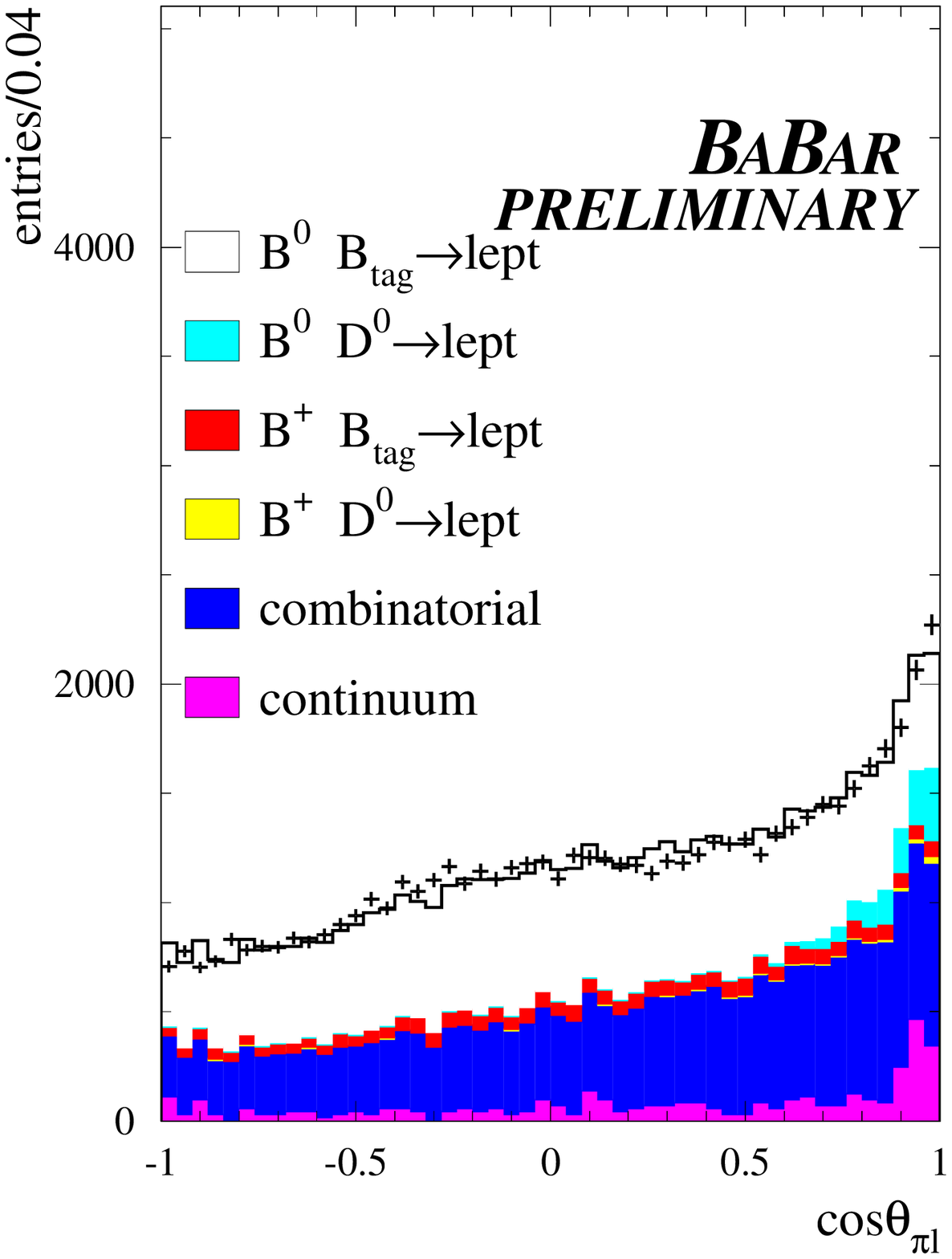}&
\hs{-2.7cm}\includegraphics[width=8cm,height=9cm]{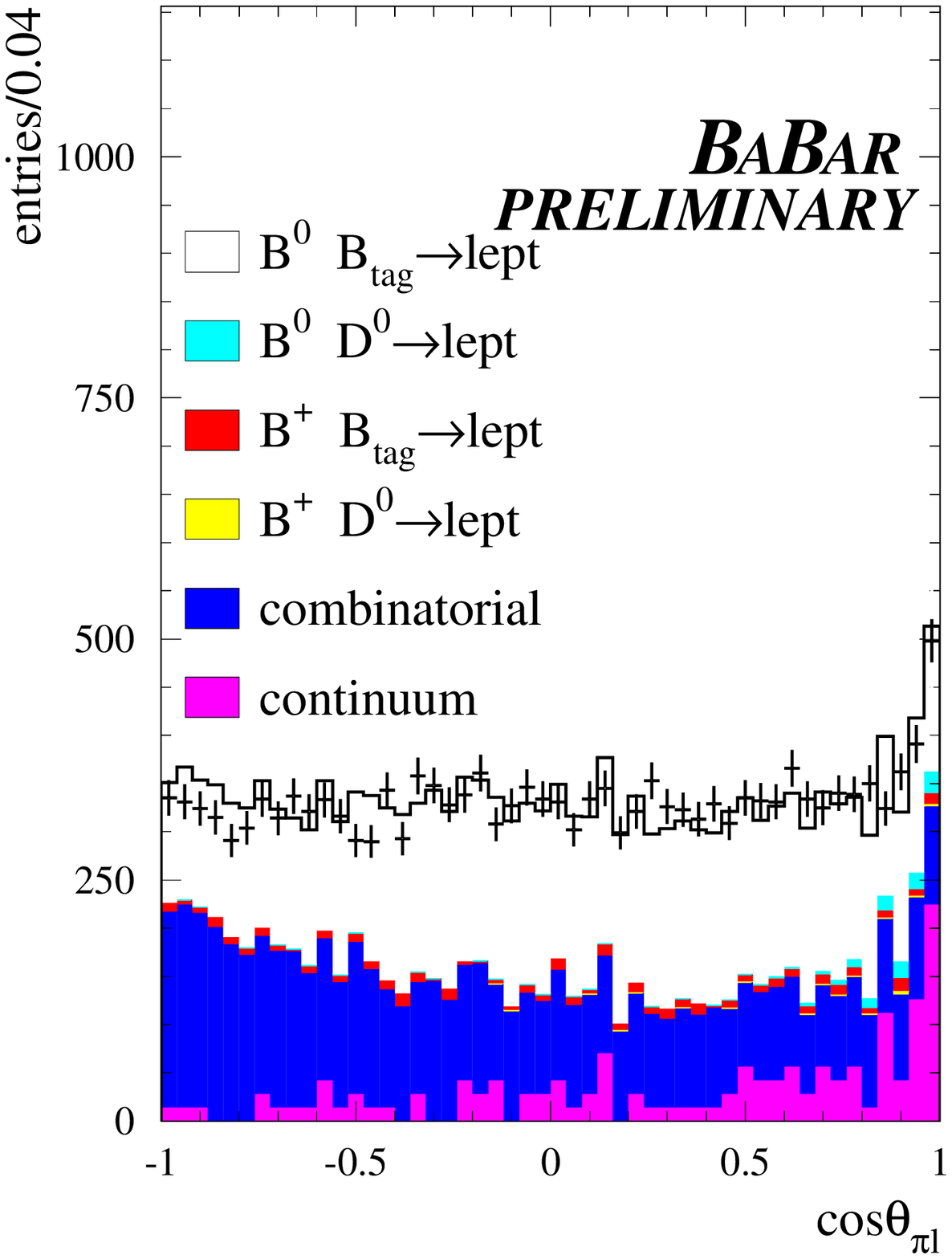}&
\hs{-2.7cm}\includegraphics[width=6cm,height=9cm]{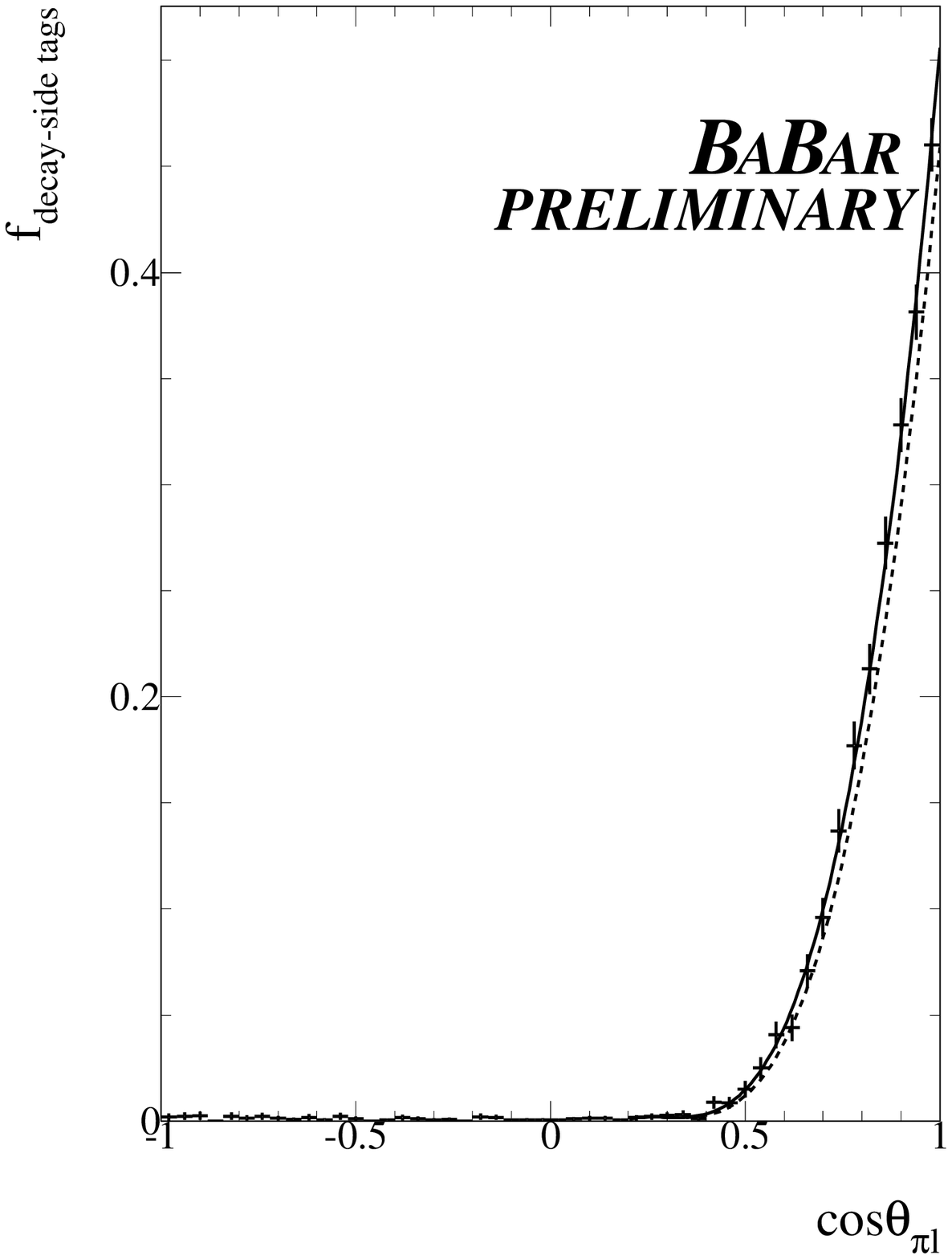}
\end{tabular} 
\end{center}
\caption{ Distribution of $\cos\theta_{\psoft \ell}$ for unmixed (left)
and mixed (center) events. The points with error bars are the data, the shaded histograms
show the various sample components after the fit. 
Right plot shows the fraction of tags from \Dz\ decay as a function
of $\cos\theta_{\psoft \ell}$ for signal events. The points with error bars are the data,
the solid line is the resulting fitted function and the dashed line is the simulation prediction.
}
\label{f:costh}
\end{figure}

Decay-side tags are produced by the semileptonic decay of the unreconstructed \Dz . Therefore
they do not carry any information about $\tBz$ or \deltamd.  The PDF  for both mixed and unmixed contributions
is a purely exponential function, with an effective lifetime representing the displacement of 
the lepton production point from the \Bzb\ decay point due to the finite lifetime of the \Dz .
We determine the fraction of these events by fitting 
the $\cos\theta_{\psoft \ell}$ distribution  (see Fig.~\ref{f:costh}),
where $\theta_{\psoft \ell}$ is the angle between the soft pion and the tag lepton. 
Using the results of the  $\cos\theta_{\psoft \ell}$ fit we parameterize the probability for each event to have a decay-side tag as a third order polynomial 
function of $\cos\theta_{\psoft \ell}$~(see Fig.~\ref{f:costh}). 
\par The signal PDF for both mixed and unmixed events consists of the sum of 
primary, cascade, and decay-side tags,
each convoluted with its own resolution function. The parameters $S_n,~S_w,~S_o,~f_w$, and 
$f_o$ are common to the three terms, while each tag type has different offsets ($o_n,o_w$).
 To reduce the systematic error, some
other parameters, in addition to $\tBz$ and \deltamd\ are free in the fit, namely,
all the parameters of the resolution functions, the dilution of the primary tags, 
the fraction of cascade tags, and the effective lifetime of the decay-side tags. We fix the
other parameters (dilution of cascade tags, fraction of decay-side tags) to their values, obtained
as described above, and then vary 
them within their error ranges to assess the corresponding systematic error, as we have described above.
\par 
We adopt a similar PDF for peaking \Bub\ background, accounting for primary, cascade, and decay terms.
Because \Bub\ mesons do not oscillate, we use a pure exponential PDF for the primary and cascade tags
with lifetime $\tBu = 1.671\pm0.018$~ps \cite{ref:HFAG}. We force the parameters of the resolution function
to equal those for the corresponding signal term. 
\par 
We describe continuum events with an exponential function convoluted with a three-Gaussian resolution
function. The mixed and
unmixed terms have separate effective lifetime parameters. All the parameters of the continuum resolution function
 are set equal to those of the signal, except for the offsets, which are free in the fit.
\par
The PDF for combinatorial \BB\ background accounts for oscillating and non-oscillating terms.
It has the same functional form as the PDF for peaking events, but with
independent parameters for the oscillation frequency and the lifetimes, which we interpret as
effective parameters that do not necessarily have a precise physical interpretation. 
The parameters  $S_n,~S_w,\mathrm{and} ~f_w$ are set to the same values as those in the signal.
\par

%% file: Res.tex
We first perform the measurement on several Monte Carlo samples.
We validate each term of the PDF by first fitting signal events, for
primary, cascade and decay-side tags separately, and then adding them together. We then
add peaking \Bub\ background, and finally add the \BB\ combinatorial background. We observe
the following features: 
\bi
\item The event selection introduces a bias of $(+0.016\pm0.005)$~ps on 
$\tBz$ and  $(+0.0027\pm0.0014)$~ps$^{-1}$ on \deltamd.
\item The boost approximation introduces a sizable bias on \deltamd\ ($- 0.0066$~ps$^{-1}$),
as observed by fitting the true \deltaz\ distribution. This bias disappears however
when we fit the smeared \deltaz\ and allow for the experimental resolution in the 
fit function.
\item Introducing \Bub peaking background does not introduce any  
additional bias.
\item Adding combinatorial \BB\ events, we introduce an additional bias of $+0.011$ \ps\ on $\tBz$
and of $-0.0058$ \ps$^{-1}$ on \deltamd.
\item All the parameters with a clear physical meaning are consistent with their expected
value; in particular the isospin scale factor ${\cal S}_{\Bub} = 1.04 \pm 0.06$ 
is consistent with unity.
\ei
Based on these observations, we correct the data results by subtracting $0.027$~ps
from $\tBz$, and adding $0.0031$~ps$^{-1}$ to $\dmd$. 
We assign 100$\%$ of the correction as a systematic uncertainty due to the observed biases.
\par
Although we determine the parameters for continuum events directly from the fit to on-peak data,
we independently fit the off-peak events to verify the consistency with the on-peak continuum
results.\par
We finally perform the fit to the on-peak data. Together with \dmd\ and $\tBz$ , we float most 
of the parameters describing the peaking \Bub , \BB\ combinatorial, and continuum background events.
The values of \dmd\ and $\tBz$ were blinded until completion of the study of the systematic  
errors. 
The unblinded fit results are
$\tBz = (1.5280 \pm 0.0084 ~\mathrm{(stat.))~ps,}$ and
$\dmd = (0.5200 \pm 0.0043 ~\mathrm{(stat.))~ps}^{-1}$.
We correct these values for the biases measured in the Monte Carlo simulation,
obtaining the preliminary \babar\ results
\ba
\label{e:res}
\tBz &=& (1.501 \pm 0.008 ~\mathrm{(stat.)} ~\pm0.030 ~\mathrm{(syst.))~ps,}\\
\nonumber \dmd &=& (0.523 \pm 0.004 ~\mathrm{(stat.)} ~\pm0.007 ~\mathrm{(syst.))~ps}^{-1}.
\ea
The correlation between \deltamd\ and $\tBz$ is $-1.2\%$. 
\deltamd\ has sizable correlations with
${\cal S}_{\Bub}$ (7.7\%) and $S_o$ (7.4\%). $\tBz$ is correlated with $f_o$ (13\%) and with a
parameter corresponding to the fraction of mixed combinatorial \Bub\ events (20\%).
The complete set of fit parameters is reported in Table~\ref{t:result}.

\begin{table}
\caption{\label{t:result}Parameters used in the PDFs from which the likelihood is calculated. 
The second column shows how
they are treated in the fit. The third (fourth) column gives the value employed in the data (MC) for the parameters that are fixed or used as a constraint.
The last column shows the sample in which the parameter is used.
The first set of parameters corresponds to peaking events, 
the second to \BB\ combinatorial events, and the third to continuum parameters. The last set refers to those parameters of the resolution function that are common to all the data sets.%The values of \dmd\ and $\tBz$ obtained in the real data are blind. 
~$\ptl$, $\ctl$ and $\cdl$ refer to primary, cascade and decay-side tags, respectively. The symbols $o$, $S$, and $f$ correspond 
to offsets, scale factors and fractions in the resolution function; the symbols $\alpha$ and $\rho$ correspond to the fraction
of decay-side tags and the fraction of mixed events in the decay-side tag sample, respectively. 
} 

\vspace{.5cm}
\bc
{\small 
\begin{tabular}{l|c|c|c|c}
Parameter       &       Usage      &  data           &  M.C.   &  Sample         \\ \hline \hline
$\tau_{B^0}$(ps)        &      free     &  1.501$\pm$0.008    & 1.575$\pm$0.006        &  \Bzb \ptl, \ctl \\
\dmd~(ps$^{-1}$) &      free     &  0.523$\pm$0.004    & 0.469$\pm$0.002        &    \Bzb \ptl, \ctl \\ 
\tDe~(ps)        &      free     &  0.16$\pm$0.02    & 0.22 $\pm$0.01        &    \Bzb, \Bub and \\
                 &               &                   &                       &     unmixed-\BB,  \cdl \\
$\tau_{\Bub}$(ps)        &       fixed      & 1.671$\pm$0.018     & 1.65 (M.C.) & \Bub and \BB   \\
${f_{\Bub}}$    & constrained   & 1.11$\pm$0.11  & 1.04$\pm$0.06          &                   \\
                & (1.0$\pm$0.5) &                &                        &                   \\               
$f_{\ctl}$      &      free     &  0.095$\pm$0.003    &  0.077$\pm$0.002       &    \Bzb           \\
$D_{\ptl}$      &    free       &  0.998$\pm$0.002   &  0.969$\pm$0.002       &    \Bzb,\ptl      \\
$D_{\ctl}$      &    fixed      &  0.65$\pm$0.08   & 0.536 & \Bzb,\ctl \\
$o_{\ptl,N} (=o_{\ptl,W})$(ps)  &      free     & $-$0.012$\pm$0.008   &  $-$0.012$\pm$0.004  &    \Bzb, \Bub \ptl \\
$o_{\ctl,N}$~(ps)           &      free     & $-$0.17$\pm$0.06     &  $-$0.52$\pm$0.04       &    \Bzb \ctl \\
$o_{\ctl,W}$~(ps)           &      free     & $-$6.2$\pm$0.8     &  $-$5.8$\pm$0.5       &    \Bzb \ctl \\
$o_{\cdl,N} (=o_{\cdl,W}$)~(ps)    &      free  &  $-$0.13$\pm$0.02   &  $-$0.13$\pm$0.01       &    \Bzb,\Bub, \cdl \\
$S_O$~(ps)          &      free     &  24.7$\pm$3.6      &  41.6$\pm$0.3       &    \Bzb, \Bub and \cdl \\
$f_O$           &      free     & 0.00004$\pm$0.00013 &  0.0013$\pm$0.0004       &    \Bzb, \Bub and \cdl  \\
\hline \hline
$\tau_{B^0}^{BKG}$~(ps)  &      free     &  1.24$\pm$0.04    &  1.26$\pm$0.07       &  \BB \\
$\Delta m_d^{BKG}$~(ps$^{-1}$)   &      free     &  0.45$\pm$0.01    &  0.50$\pm$0.02       &  \BB \\ 
\tDe$_{mixed}^{BKG}$~(ps) &      free     &  2.4$\pm$0.5      &  1.2$\pm$1.7       &  \BB, \cdl\ (mixed only) \\
${f^{\Bub}}$         &      free     &  0.63$\pm$0.01      &  0.60$\pm$0.02       &  \BB   \\
${g^{\Bub}}$         &      free     &  0.043$\pm$0.001      &    0.078$\pm$0.003    &  \BB   \\
$f_{bcl}^{BKG}$      &      free     &  0.0001$\pm$0.0001      &    0.01$\pm$0.01    &  \BB  \\
$D_{\ptl}^{BKG}$     &      free     &  0.997$\pm$0.003      &    0.986$\pm$0.004    &  \BB  (\Bzb\ only)       \\
$o_{BKG,\ptl}$~(ps)      &      free     &  $-$0.06$\pm$0.03      &   $-$0.07$\pm$0.12   &  \BB \\
$o_{BKG,\ctl}$~(ps)       &      free     &  $-$16.7$\pm$2.1      &  $-$1.8$\pm$0.4       &  \BB \\
$o_{BKG,\cdl}$~(ps)       &      free     &  0.01$\pm$0.02   &  $-$0.06$\pm$0.01       &  \BB, \cdl  \\
$\alpha_{B0}^{BKG}$  &      free     &  0.28$\pm$0.02  &  0.29$\pm$0.04       &  \BB (\Bzb only) \\
$\rho_{B0}^{BKG}$    &      free     &  0.0$\pm$0.1    &  0.2$\pm$0.1       &  \BB (\Bzb only) \\
$\alpha_{B+}^{BKG}$  &      free     &  0.039$\pm$0.004 &  0.134$\pm$0.006       &  \BB (\Bub only) \\
$\rho_{B+}^{BKG}$    &      free     &  0.03$\pm$0.11      &    0.2$\pm$0.1    &  \BB (\Bub only) \\ 
$S_O^{BKG}$~(ps)      &      free     &  40.2$\pm$0.3     &  48$\pm$1      & \BB  \\
$f_O^{BKG}$          &      free     &  0.0046$\pm$0.0014 &  0.0011$\pm$0.0007       & \BB  \\ \hline \hline
$\tau_{lq}$~(ps)          &      free     &  0.17$\pm$0.02    &    -     &  Continuum \\
$o_{lq,N}=o_{lq,W}$~(ps)  &      free     &  $-$0.05$\pm$0.01   &    -     &  Continuum \\ \hline \hline
$S_N$           &      free     &       0.997$\pm$0.006     &  0.996$\pm$0.004       &    common to all  \\
$S_W$           &      free     &       3.74$\pm$0.14       &  2.48$\pm$0.06       &    common to all  \\
$f_W$           &      free     &       0.024$\pm$0.001   &  0.036$\pm$0.002       &    common to all  \\
$o_O$~(ps)      &      fixed      &   0     &    0    &    common to all  \\ \hline \hline
$\rho$($\tBz$,\dmd)& &$-$0.012 &$-$0.081  \\ \hline \hline
\end{tabular} }
\ec 
\et
Details on the systematic error are reported in Section \ref{sec:Systematics}. Figure~\ref{f:fitmass}
shows the comparison between the data and the fit function projected on \deltat ,
for a sample of events enriched in signal by the cut  $\mnusq > -2.5$~GeV$^2/c^4$; 
Fig.~\ref{f:fitside} shows the same comparison for events in the background region.
Figure~\ref{f:asy} shows the plot of the time-dependent asymmetry 
${\cal A}(\deltat)=\frac{N_{Unmixed}(\deltat)-N_{Mixed}(\deltat)}{N_{Unmixed}(\deltat)+N_{Mixed}(\deltat)}$
for events in the signal region and events in the background region.\par
The agreement between the fitting function and the data distribution is good both in the signal
and in the background regions. We perform a set of parametrized Monte Carlo experiments to assess 
the quality of the fit.
The probability to obtain a lower likelihood is 50$\%$.
\begin{figure}[!htb]
\begin{center}
\begin{tabular}{cc}
\hs{-1.5cm}\includegraphics[width=9cm,height=9cm]{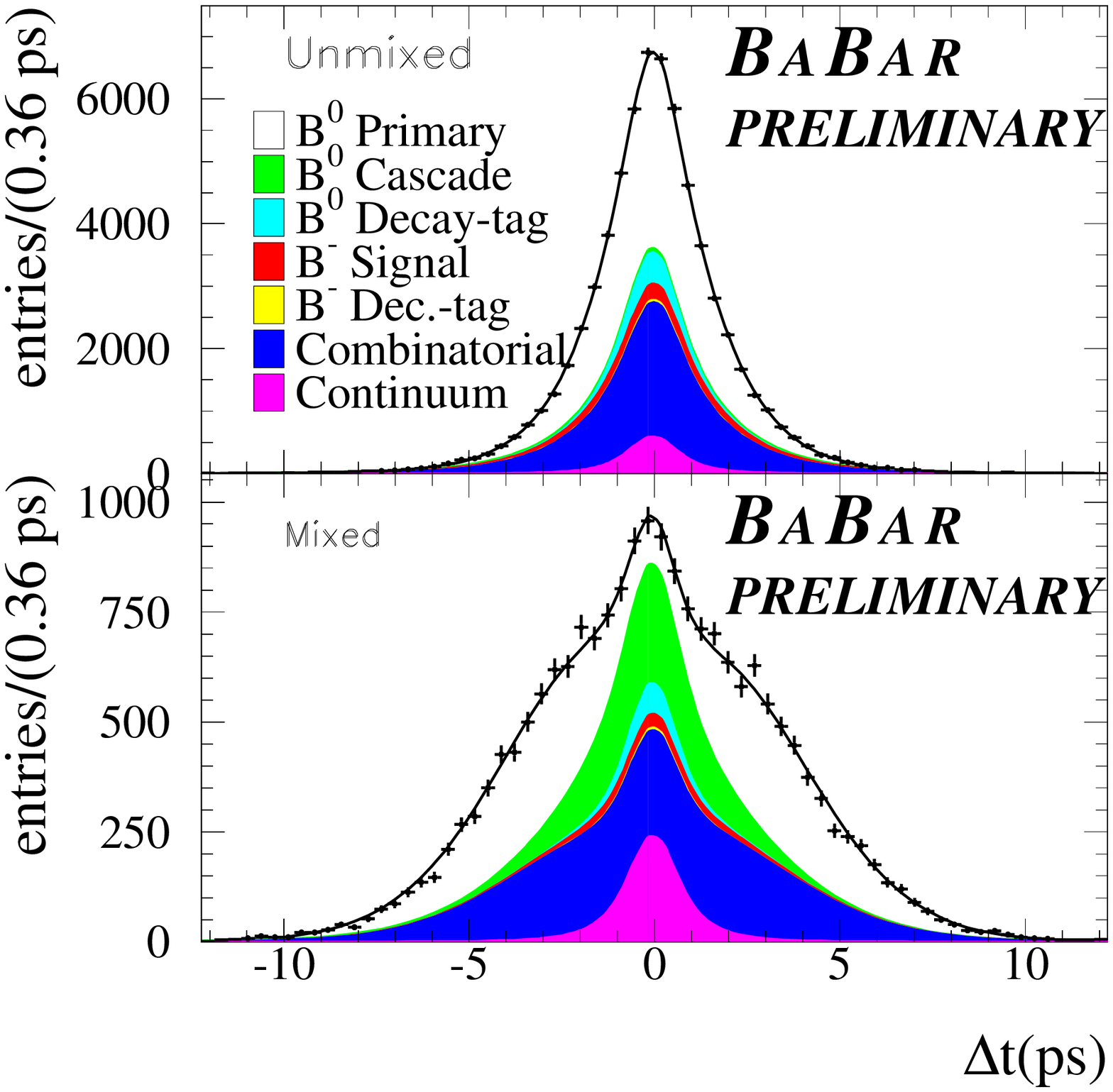}&
\hs{-.5cm}\includegraphics[width=9cm,height=9cm]{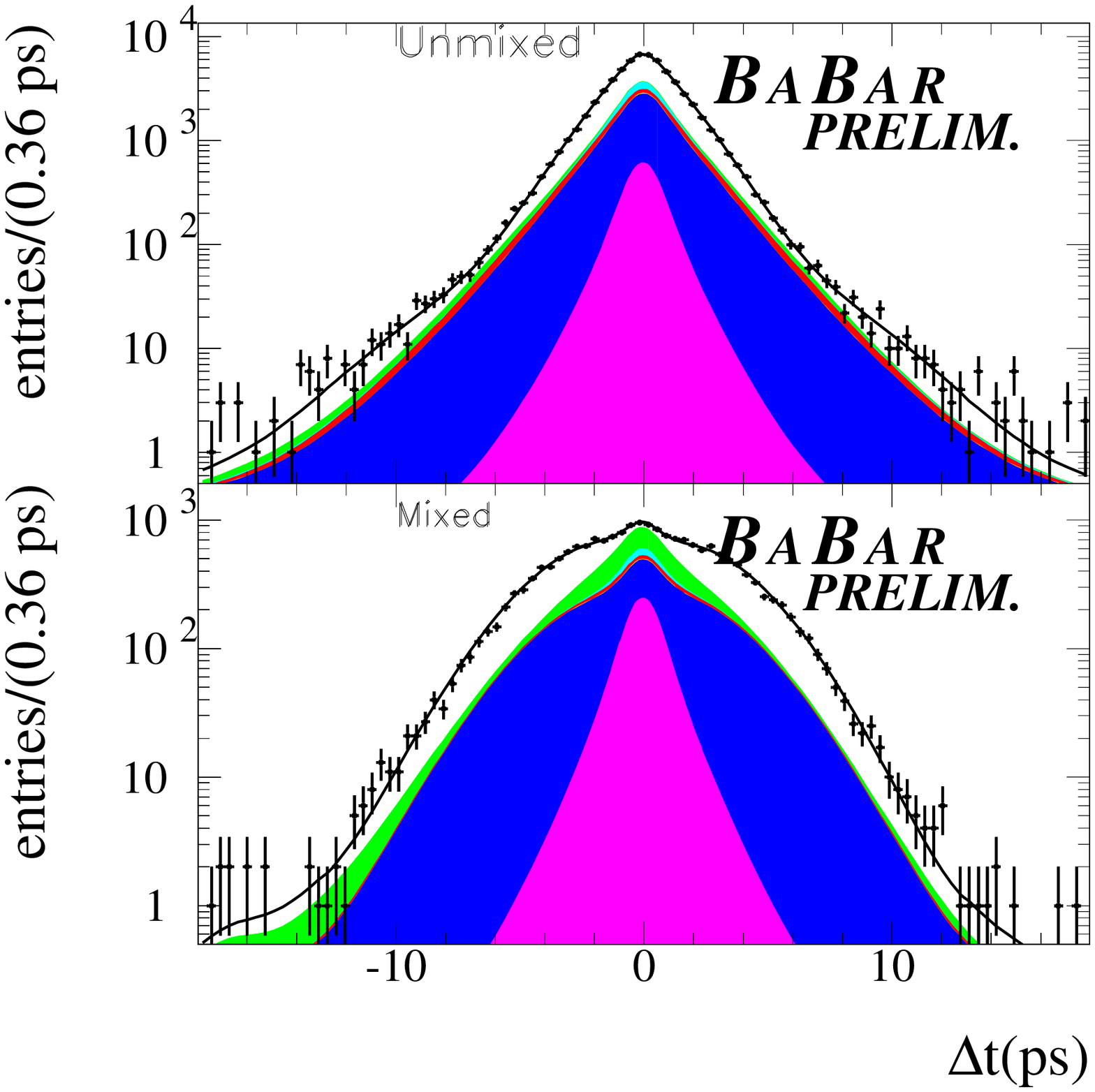}
\end{tabular} 
\end{center}
\caption{Distribution of \deltat\ for unmixed (top) and mixed (bottom) events in the signal $\mnusq$
region with linear (left) and logarithmic (right) scales. The points show the data, the curve is 
the projection of the fit result,
and the shaded areas from bottom to top are the contributions from continuum, \BB\ combinatorial, peaking \Bub\ with decay-side tag, peaking \Bub\ with primary tag, 
signal with decay-side tag, signal with cascade tag, and signal with primary tag. } 
\label{f:fitmass}
\end{figure}
\begin{figure}[!htb]
\begin{center}
\begin{tabular}{cc}
\hs{-1.5cm}\includegraphics[width=9cm,height=9cm]{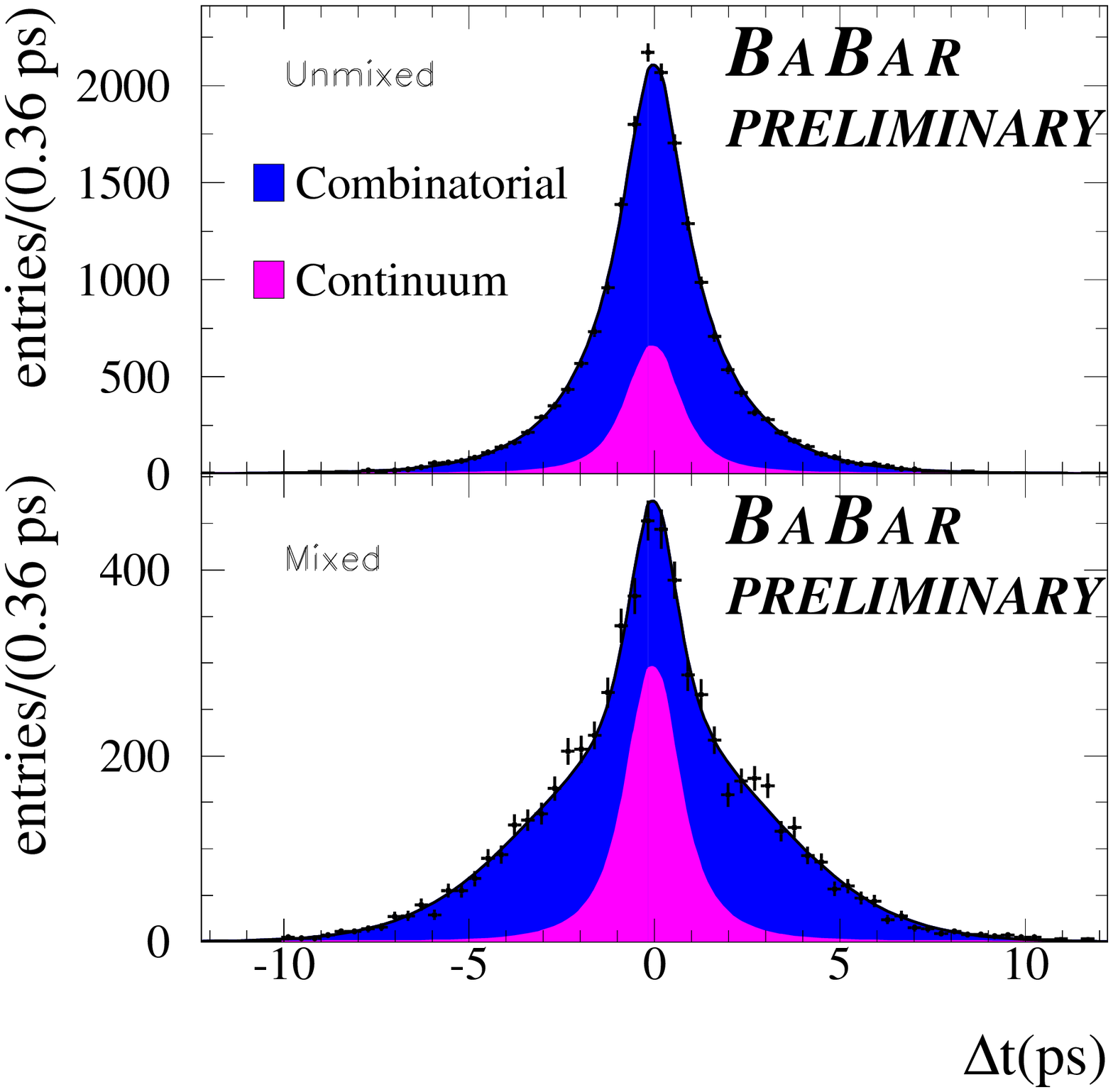} &
\hs{-.5cm}\includegraphics[width=9cm,height=9cm]{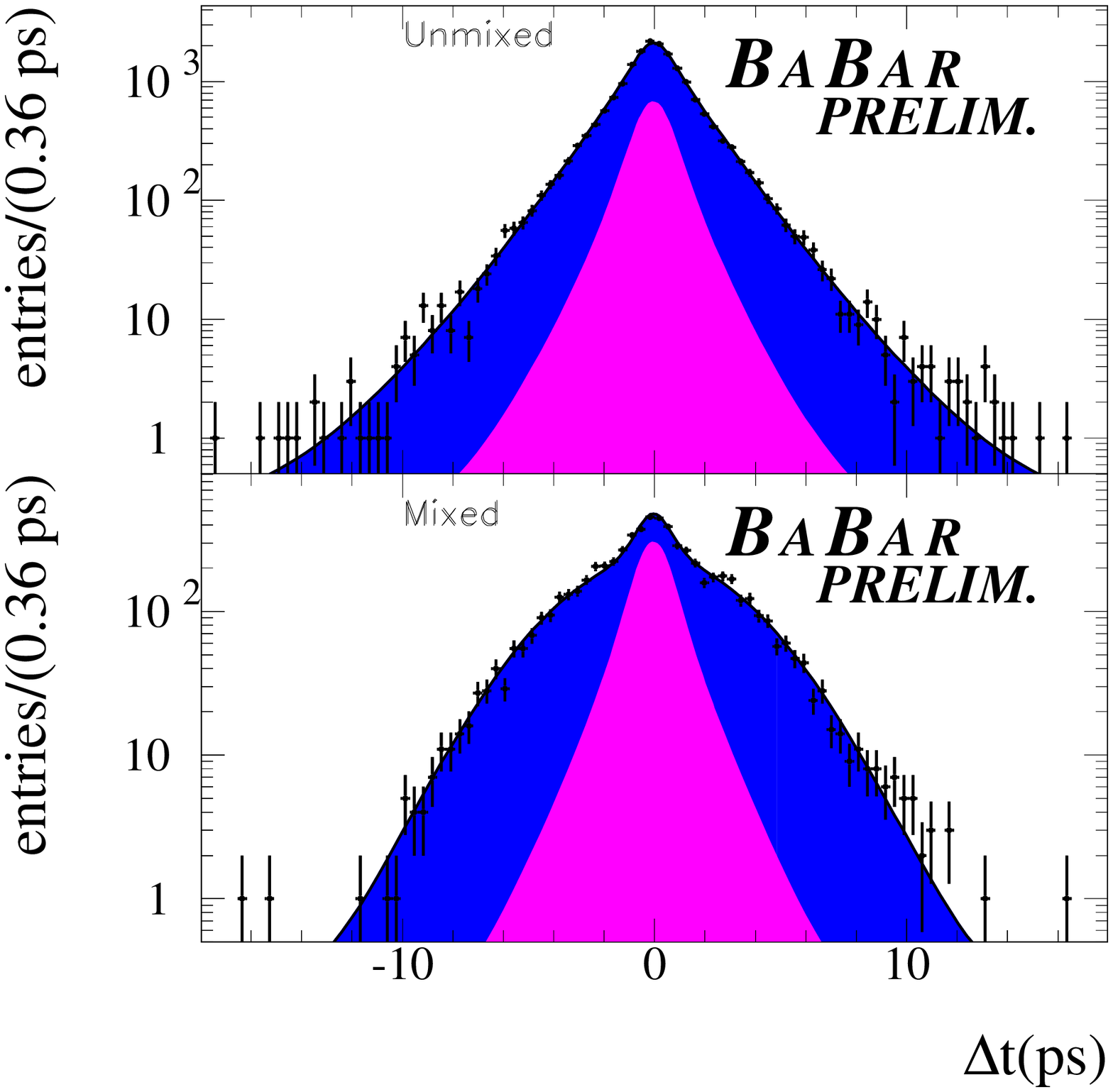}
\end{tabular} 
\end{center} 
\caption{Distribution of \deltat\ for unmixed (top) and mixed (bottom) events in the background 
$\mnusq$
region with linear (left) and logarithmic (right) scales. The points show the data, the curve is 
the projection of the fit result, and
the shaded areas are the contributions from continuum and \BB\ combinatorial background.  } 
\label{f:fitside}
\end{figure}

\begin{figure}[!htb]
\begin{center}
\begin{tabular}{cc}
\hs{-1.5cm}\includegraphics[width=9cm,height=9cm]{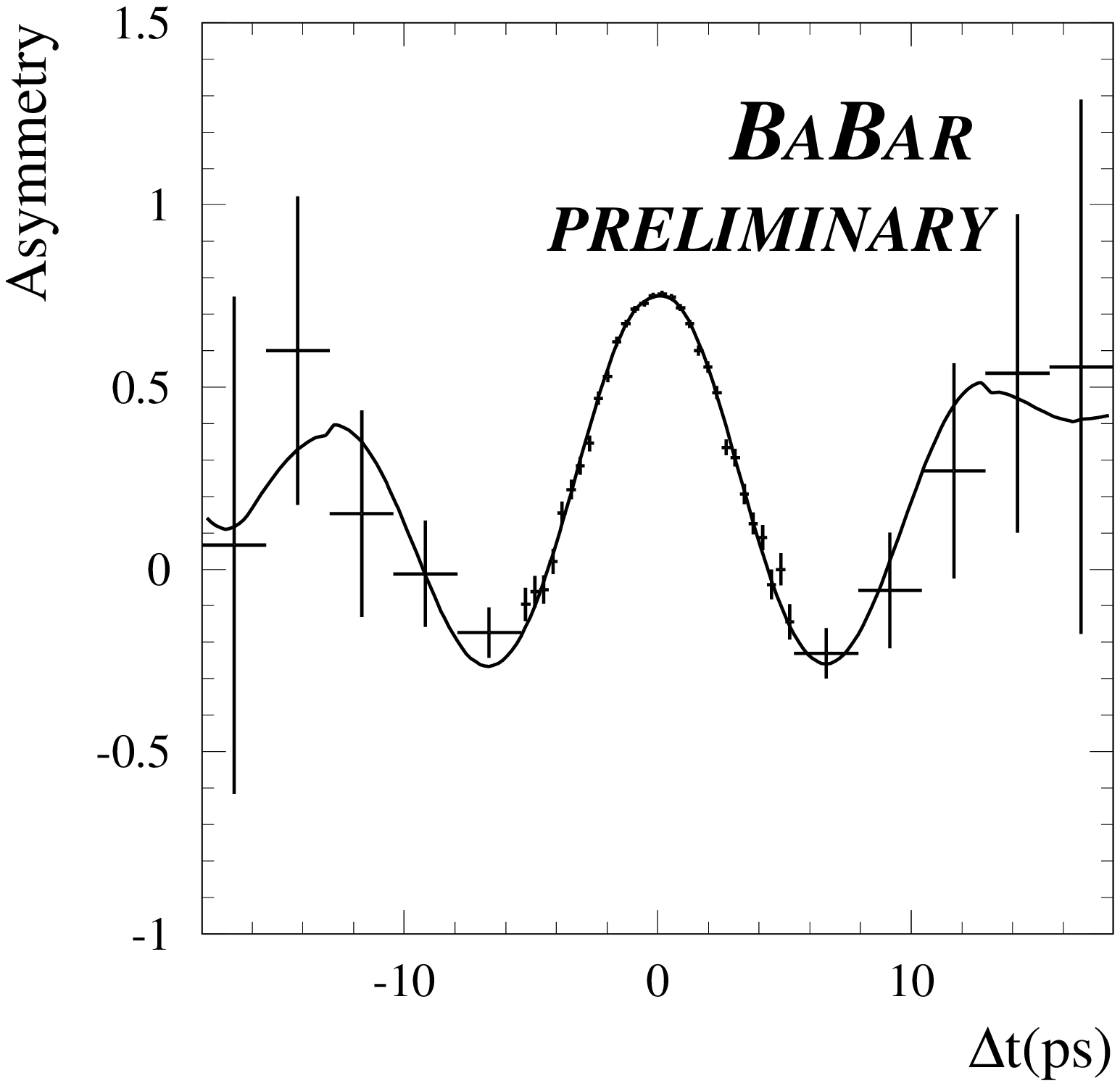} &
\hs{-.5cm}\includegraphics[width=9cm,height=9cm]{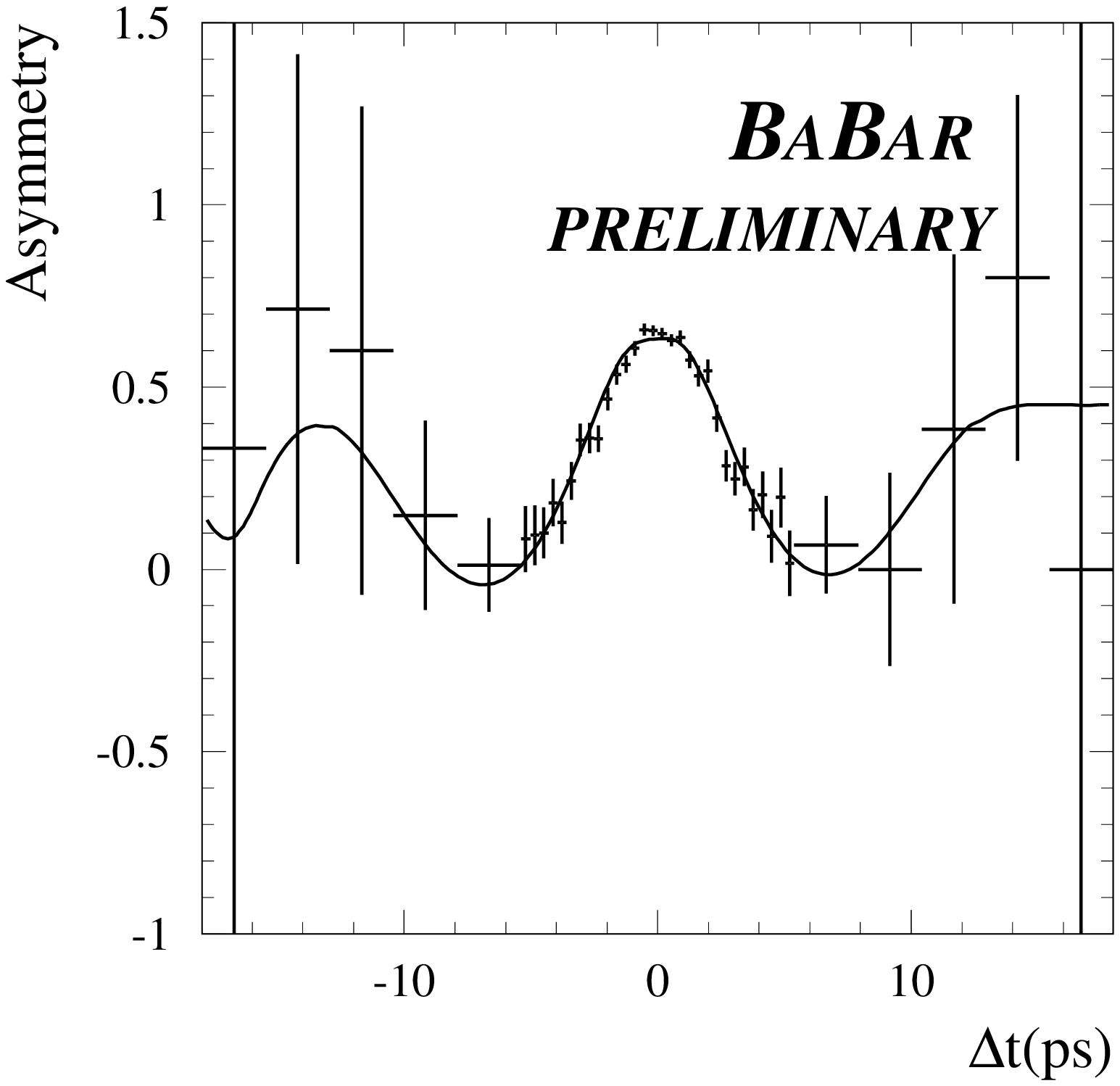}
\end{tabular} 
\end{center}
\caption{Asymmetry between unmixed and mixed events as a function of \deltat,
for events in the signal region (left) and in the background region (right). Points with error bars are the
data, and the curve is a projection of the fit result.}
\label{f:asy}
\end{figure}

%% file: Syst.tex
The systematic errors are summarized in Table~\ref{t:syst}.
We consider the following sources of systematic uncertainty:
\ben
\item Sample composition: the fraction of peaking \Bub\ events in the 
sample, ${\cal S}_{\Bub}$, is free in the fit; therefore we do not assign any 
systematic error on this fraction. As mentioned in Section~\ref{s:sample}, 
there is a $\pm 2.3 \%$ systematic uncertainty on the background rate  
due to possibly peaking combinatorial background. We therefore vary the fraction of \BB\
events in the signal region by that amount, repeat the fit, and add the variation in the result
to the systematic error
(entry (a) of Table \ref{t:syst}). We neglect the statistical error on the sample composition because it 
is significantly smaller than  this systematic effect.
\item Analysis bias (entry b): we take 100$\%$ of the bias observed in the fit on the Monte Carlo sample.
\item Signal and background PDF description: most of the parameters in the PDF are free in the fit
and therefore do not contribute to the systematic error. We vary the parameters that are fixed in the 
fit by their uncertainty, repeat the fit, and take the corresponding variation in $\tBz$ and \dmd\ as systematic errors.
We take the uncertainty on \tBu (entry (c)), and on $D_{\ctl}$ (entry (d)) from the PDG~\cite{ref:PDG}. 
\item We consider effects due to the detector $z$ scale (entry (e)), the knowledge of the PEP-II boost 
(entry (f)), the actual position of the beam spot (entry (g)), and SVT alignment (entry (h)). 
Detailed studies of these effects have been 
performed in \babar\ in other mixing and lifetime analyses 
(dilepton \cite{ref:dilepton} and fully reconstructed 
\BtoDs\ ~\cite{ref:xl} analyses) and provided consistent results. 
As the methods for vertex reconstruction are 
very similar we provisionally take 
our systematic error due to these effects from Ref.~\cite{ref:xl}.
\item We vary the parameters describing the fraction of decay-side tags by their statistical errors (entry (i)).
\item Binned fitting: we vary the number of bins in \deltat\ from 100 to 250 
and in \st\ from 20 to 50, and we repeat the fit. Alternatively, we use the average value of the 
likelihood in the bin instead of the value corresponding to the center of the bin. We take the
systematic error to be the maximum variation with reference to the default result (entry (j)).
\item Outlier description: we vary the value of the offset of the outlier Gaussian  
from $-5~{\rm ps}$  to $5~{\rm ps}$. Alternatively, we use a flat PDF for their description (entry (k)).
\item Fit range: we vary the \deltat\ fit range from $\pm$ 18~{\rm ps} to
$\pm$ 10~{\rm ps} and the \st\ range between 1.8~{\rm ps} and 4.2~{\rm ps} (entry (l)).

\bt \bc 
\caption{\label{t:syst} Systematic uncertainties.}
\begin{tabular}{|l|c|c|c|c|}
\hline
Source  & Variation & $\delta\tBz$ (ps) & $\delta\dmd$ (ps$^{-1}$) \\ \hline \hline
                             &                &                &               \\
(a) \BB\ fraction            &  $\pm 2.3\%$   &   $\pm0.001$  & $\pm0.001$   \\
(b) Analysis bias            &       -        &   $\pm0.027$  & $\pm0.003$   \\
(c) \tBu\                    & 1.671$\pm$0.018 ps&   $\pm0.002$  & $\pm0.001$   \\  
(d) ${\cal D_{\ctl}}$        & 0.65$\pm$0.08  &   $\pm0.005$  & $\pm0.001$  \\ 
(e) $z$ scale                &     -          &   $\pm 0.006$  &  $\pm0.002$     \\ 
(f) PEP-II boost             &     -          &   $\pm 0.002$  &  $\pm0.001$     \\ 
(g) Alignment                &     -          &   $\pm 0.006$  &  $\pm0.003$     \\ 
(h) Beam spot position       &     -          &   $\pm 0.005$  &  $\pm0.001$     \\  
(i) Decay-side tags          &     -          &   $\pm 0.002 $ &  $\pm0.001$     \\ 
(j) Binning                  &     -          &   $\pm 0.002 $ &  $\pm0.002$     \\ 
(k) Outlier                  &     -          &   $\pm 0.001 $ &  $\pm0.002$     \\ 
(l) \deltat\ and \st\ cut    &     -          &   $\pm 0.008 $ &  $\pm0.003$     \\ 
\hline
 Total                       &                &   $\pm 0.030$  &  $\pm0.007$    \\ 
 \hline \hline

\end{tabular}
\ec \et
\een

\section{CONSISTENCY CHECKS}
We rely on the assumption that the parameters of the background PDF do not depend on $\mnusq$.
We verify this assumption for the continuum background with the fit to the off-peak events. 
To check this assumption for the \BB\ combinatorial PDF, we perform several cross checks on the data and the Monte Carlo.
We compare the simulated \BB \deltat\ distribution in several independent regions of $\mnusq$ 
with Kolmogorov-Smirnov tests and always obtain a reasonable probability for agreement. We fit \BB\ events 
separately in the signal and background $\mnusq$ region and compare the parameters of the PDF.
We fit the signal plus background Monte Carlo events in the signal region only,
fixing all the parameters of the \BB\ sample to the values obtained in a fit in the background region, and do not
see any appreciable deviation from the result of the full fit.
Finally, we repeat 
the fit both on the data and the Monte Carlo using different $\mnusq$ ranges for the background region.
Once  again, we do not observe any appreciable difference in $\tBz$ and \dmd\ relative to the default result.

%% file: Conclusion.tex
% ====================
%\section{Conclusion} 
\label{Conclusion}
% ====================
We have performed a measurement of \dmd\ and $\tBz$ on a sample of
50000 partially reconstructed, lepton-tagged \BtoDs\ decays. We obtain the
following preliminary results:
\ba
\nonumber \tBz &=& (1.501  \pm 0.008  ~\mathrm{(stat.)} ~\pm0.030  ~\mathrm{(syst.))~ps},\\
\nonumber \dmd &=& (0.523 \pm 0.004 ~\mathrm{(stat.)} ~\pm0.007 ~\mathrm{(syst.))~ps}^{-1}.
\ea
These preliminary values are consistent with published measurements of \tBz\ and \dmd\ 
performed by \babar\ with different data sets, and with the world averages computed
by the Heavy Flavor Averaging Group \cite{ref:HFAG}, as can be seen in 
table \ref{t:avr}. The error we obtain on \dmd\ is comparable to the 
uncertainty on the present world average.

\bt
\caption{ \label{t:avr}Comparison of this result with previous \babar\ measurements and
 with the world average. In the case of the world average, we report the
total error; in all other cases we show the statistical and systematic errors
separately. P.R. means partial reconstruction.}
\bc\begin{tabular}{l|c|c|c}

\hline \hline
Method  & \dmd\ (ps)$^{-1}$ & \tBz\ (ps) & Reference \\
\hline
\babar\ Hadronic   & $0.516 \pm 0.016 \pm 0.010$ & $1.546\pm0.032 \pm 0.022 $ & \cite{ref:hdmd,ref:htau}\\
\babar\ $\ell\ell$ & $0.493 \pm 0.012 \pm 0.009$ &  $-$ & \cite{ref:dilepton}  \\
\babar\ $\dsm \ellp \nul$  & $0.492 \pm 0.018 \pm 0.013$ &  $1.523 \pm 0.024 \pm 0.022$ & \cite{ref:xl}\\
\babar\ $\dsm \ellp \nul $(P.R.)  & $-$ &  $1.529 \pm 0.012 \pm 0.029$ & \cite{ref:t1} \\
\babar\ $\dsm \pi^+$(P.R.)  & $-$ &  $1.533 \pm 0.034 \pm 0.038$ & \cite{ref:t2} \\ 
World Average & $0.502\pm0.007$ & $1.536\pm0.014$ & \cite{ref:HFAG} \\ \hline
This Measurement & $0.523 \pm 0.004 \pm 0.007$ & $1.501  \pm 0.008 \pm 0.030$ & \\ \hline
\end{tabular}\ec\et

%% file: acknowledgements.tex
We are grateful for the 
extraordinary contributions of our \pep2\ colleagues in
achieving the excellent luminosity and machine conditions
that have made this work possible.
The success of this project also relies critically on the 
expertise and dedication of the computing organizations that 
support \babar.
The collaborating institutions wish to thank 
SLAC for its support and the kind hospitality extended to them. 
This work is supported by the
US Department of Energy
and National Science Foundation, the
Natural Sciences and Engineering Research Council (Canada),
Institute of High Energy Physics (China), the
Commissariat \`a l'Energie Atomique and
Institut National de Physique Nucl\'eaire et de Physique des Particules
(France), the
Bundesministerium f\"ur Bildung und Forschung and
Deutsche Forschungsgemeinschaft
(Germany), the
Istituto Nazionale di Fisica Nucleare (Italy),
the Foundation for Fundamental Research on Matter (The Netherlands),
the Research Council of Norway, the
Ministry of Science and Technology of the Russian Federation, and the
Particle Physics and Astronomy Research Council (United Kingdom). 
Individuals have received support from 
CONACyT (Mexico),
the A. P. Sloan Foundation, 
the Research Corporation,
and the Alexander von Humboldt Foundation.